\documentclass{article} \usepackage{slashed,feynmf,epsfig,amsmath,amssymb,enumitem} \usepackage[papersize={8.5in,11in}]{geometry}
   \usepackage{bm}
\begin{document} 
\title{Inhomogeneous Cosmology Redux} 
\author{J. W. Moffat\\~\\
Perimeter Institute for Theoretical Physics, Waterloo, Ontario N2L 2Y5, Canada\\
and\\
Department of Physics and Astronomy, University of Waterloo, Waterloo,\\
Ontario N2L 3G1, Canada}
\maketitle



\begin{abstract}
An alternative to the postulate of dark energy required to explain the accelerated expansion of the universe is to adopt an inhomogeneous cosmological model to explain the supernovae data without dark energy. We adopt a void cosmology model, based on the inhomogeneous Lema\^{i}tre-Tolman-Bondi solution of Einstein's field equations. The model can resolve observational anomalies in the $\Lambda CDM$ model, such as the discrepancy between the locally measured value of the Hubble constant, $H_0=73.24\pm 1.74\,{\rm km}\,{\rm s}^{-1}\,{\rm Mpc}^{-1}$, and the $H_0=66.93\pm 0.62\,{\rm km}\,{\rm s}^{-1}\,{\rm Mpc}^{-1}$ determined by the Planck satellite data and the $\Lambda CDM$ model, and the lithium $^{7}{\rm Li}$ problem, which is a $5\sigma$ mismatch between the theoretical prediction for the $^{7}{\rm Li}$ from big bang nucleosynthesis and the value that we observe locally today at $z=0$. The void model can also resolve the tension between the number of massive clusters derived from the Sunyaev-Zel'dovich effect by the Planck satellite and the number expected from the CMB anisotropies, and the CMB weak lensing anomaly observed in the Planck data. The cosmological Copernican principle and the time and position today coincidence conundrums in the $\Lambda CDM$ and void cosmological models are discussed.
\end{abstract}

\maketitle


\section{Introduction}

The standard model of cosmology - the $\Lambda CDM$ model - is remarkably successful in fitting the large range of observations in spite of its simplicity and a handful of parameters. The model's central postulate is the cosmological Copernican principle, that we are not at a special time or spatial location in the universe. Related to this principle is the Cosmological principle: Smoothed on a large enough scale the universe is spatially homogeneous and isotropic. Adopting the maximally symmetric, homogeneous and isotropic Friedmann-Lema\^{i}tre-Robertson-Walker (FLRW) spacetime with constant spatial curvature as a background, Newtonian perturbative calculations suffice to produce the main predictions of the model~\cite{Buchert}. Yet the physical nature of dark matter and the accelerated expansion of the Universe~\cite{Perlmutter,Riess} are not understood at a fundamental level. The cold dark matter required to explain the CMB data and the growth of structure in the early universe has not been so-far detected in the present late-time Universe~\cite{LUX,PANDAX}. The dark energy that drives the accelerated expansion of the Universe is not understood. The cosmological constant plays the role of dark energy at the price of introducing a severe fine-tuning problem. The cosmological constant can be explained as a quantum vacuum energy with an equation of state corresponding to a fluid with negative pressure, in a way that corresponds to the equation of state for the vacuum energy in particle physics. But there is a huge discrepancy between the value of the energy density estimated from cosmological observations and the energy density determined by quantum theory of order $10^{122}$~\cite{Weinberg}. Modified gravity theories, which introduce new degrees of freedom, have been proposed to explain the dark energy~\cite{Capoziello}.

A geometrical spacetime description of cosmology called the Void Cosmology~\cite{Moffat1,Moffat2,Clarkson,Moffat3,Moffat4,Moffat5}, adopts the idea that we live in a large void embedded in an asymptotic LFRW spacetime. The model is based on the inhomogeneous Lema\^{i}tre-Tolman-Bondi (LTB) exact, pressureless solution of Einstein's field equations~\cite{Lemaitre,Tolman,Bondi}. Inhomogeneous cosmological models with more general spacetime geometries have been proposed~\cite{Szekeres,Szafron,Wainright,Krasinski}. The spatial homogeneity of the universe is not yet established independently of the standard model paradigm including dark energy~\cite{Bull}. 

The void cosmology model can explain the luminosity distance with a spatially varying energy density, Hubble parameter and curvature on Gpc scales. The cosmological constant is zero and there is no acceleration of the expansion of the universe; the dimming of light rays from supernovae is due to the light rays passing through the void, making the supernovae appear to be more distant than they would be in a region of spacetime with a higher density of matter. The model assumes a spherically symmetric spacetime and to avoid excessive anisotropy and cosmic dipole moment the observer on earth has to be within tens of Mpc of the center of the spherically symmetric void, which implies a spatial coincidence of about $(40\, {\rm Mpc}/15\, {\rm Gpc})^3\sim 10^{-8}$. 

This anti-Copernican feature of a simple void cosmology has to be balanced against the severe fine-tuned coincidence problem in the standard $\Lambda CDM$ model; the values of $\Omega_M$ and $\Omega_\Lambda$ are nearly equal now, though they were not in the past and will not be in the future. This is related to the problem of time synchronicity that combined cosmological and astrophysical data yield for the the present value of the dimensionless age of the universe, $H_0t_0=0.96\pm 0.01$~\cite{Tonry,Kirshner}. This temporal coincidence problem occurs even though the universe was decelerating for its first 9 Gyr and then suffered a cosmic acceleration for the past 5 Gyr, and it is particularly strange because over the span of time, the dimensionless age of the universe can take on a wide range of values $0 < H_0t(a(t)) < \infty$, where $a(t)$ is the cosmic scale. The relation $H_0t_0=1$ holds exactly in the Milne cosmology~\cite{Milne}. When matter is introduce into the Milne cosmology, $H_0t_0 < 1$. In the $\Lambda CDM$ homogeneous model this is adjusted to $H_0t_0\sim 1$ when dark energy is included in the model. 

If we are forced observationally to the conclusion that we are at a special time today in the evolution of the universe, then we could also suppose {\it inter alia} that we are in a special position in the universe, thereby, violating the Copernican principle both in space and in time~\footnote[1]{From a philosophical point of view, this violation of the Copernican principle is not only radical but unattractive. However, only experimental observations of the universe can ultimately decide which philosophical point of view should be dominant.}. Can it be that the Copernican principle is violated at a fundamental level? The Copernican principle is hard to test on large scales, because we view the universe from one spacetime event, although attempts can be made to test the principle locally~\cite{Ellis,ClarksonBassettLu}. We can place limits on anisotropy around galaxy clusters and measure bulk flow through the kinetic Sunyaev-Zeldovich (kSZ) effect~\cite{Clarkson,SZ,Bull,Goodman,Labini,Stebbins}, and possibly conceive of ways to test the observational consistency of the FLRW model in the past lightcone. 

Constraints on the Hubble constant $H_0$ coming from the Planck satellite have been in tension with the results of Riess et al.,~\cite{Riess11}, based on the Hubble Space Telescope measurements. The tension was confirmed in the Planck 2015 data release~{\cite{Planck1}. The recent results by Riess et al.,~\cite{Riess16} confirm again the tension between the local measurement of $H_0$ and the CMB Planck 2016 data~\cite{Planck3,BOSS}. The new Riess et al., data yield, $H_0=73.24\pm 1.74\,{\rm km}\,{\rm s}^{-1}\,{\rm Mpc}^{-1}$, while the Planck data and the $\Lambda CDM$ model with 3 neutrino flavors having a mass of 0.06 eV give $H_0=66.93\pm 0.62\,{\rm km}\,{\rm s}^{-1}\,{\rm Mpc}^{-1}$. The local value of $H_0$ is higher than the $\Lambda CDM$ value by 3.4$\sigma$~\cite{Riess16}~\footnote[2]{The discrepancy reduces to 2.1$\sigma$ relative to the prediction of $69.3\pm 0.7\,{\rm km}\,s^{-1}\,{\rm Mpc}^{-1}$ based on the comparably precise combination of WMAP+ACT+SPT+BAO observations. This suggests that systematic uncertainties in the CMB radiation measurements may play a role in the tension.}.

The amplitude of density fluctuations measured locally is in conflict with the predictions from the Planck CMB data~\cite{Planck2,Macaulay,BattyeMoss,Wyman,BattyeCharnock,Salvatelli,MacCrann,Liu}. The number of massive clusters obtained from the Sunyaev-Zel'dovich (SZ) effect measured by the Planck collaboration is about half that expected from the CMB anisotropies. There is a discrepancy between the amplitude of matter density fluctuations inferred from the SZ effect cluster number counts and the polarization anisotropies and the primary temperature of the CMB measured by the Planck satellite.

There is also the long-standing tension caused by the lithium $^{7}{\rm Li}$ problem. In the $\Lambda CDM$ model, the observed abundances lead to an excellent agreement for D/H at the measured value for $\eta=n_b/n_\gamma=10^{-10}\eta_{10}$. Moreover, there is no discrepancy between theory and observation for $^{4}He$. However, there is general agreement that there is a problem concerning the abundance of $^{7}{\rm Li}$~\cite{Cyburt}. For the measured value of $\eta$, the CMB predicted abundance of $^{7}{\rm Li}$ derived from the $\Lambda CDM$ model disagrees with the observationally determined value at $z=0$ by $\sim 5\sigma$. 

Many proposals have been forwarded to resolve some of these discrepancies with the standard $\Lambda CDM$ model, including massive neutrinos, decaying dark matter, extra neutrinos and phantom dark energy~\cite{Silk}. Moreover, phenomenological voids have been investigated to resolve the issue of the locally measured density fluctuations and the $H_0$ tension~\cite{Lee,Ichiki,Marra}. 

In the following, as an example of an inhomogeneous model, we will utilize a large void cosmology model to resolve the anomalous observational discrepancies with the standard $\Lambda CDM$ model. 

\section{Void Cosmology}

Consider a spherically symmetric inhomogeneous universe filled with dust. The Lema\^{i}tre-Tolman-Bondi line
element in comoving coordinates can be written as~\cite{Lemaitre,Tolman,Bondi}:
\begin{equation} 
\label{linel}
 ds^{2}=dt^{2}-R^{\prime 2}(t,r)f^{-2}dr^{2}-R^{2}(t,r)d\Omega^{2},
\end{equation}
where $f$ is an arbitrary function of $r$ only. The energy-momentum tensor ${T^\mu}_\nu$ takes the barytropic form:
\begin{equation}
\label{energymomentum} {T^\mu}_\nu=(\rho+p)u^\mu u_\nu
-p{\delta^\mu}_\nu,
\end{equation}
where $u^\mu=dx^\mu/ds$ and, in general, the density $\rho=\rho(r,t)$ and the pressure $p=p(r,t)$ depend on both $r$
and $t$. We have for comoving coordinates $u^0=1, u^i=0,\,(i=1,2,3)$ and $g^{\mu\nu}u_\mu u_\nu=1$.
 
The Einstein field equations demand that $R(t,r)$ satisfies:
\begin{equation} 
\label{F}
 2R\dot{R}^{2}+2R(1-f^{2})=F(r),
\end{equation}
where $\dot R=dR/dt$ and $F$ is an arbitrary function of class $C^{2}$. The proper density can be expressed as
\begin{equation} 
\label{dens}
 \rho=\frac{F^{\prime}}{16\pi R^{\prime} R^{2}},
\end{equation}
where $R^{\prime}=dR/dr$.

The total mass within comoving radius $r$ is given by
\begin{equation} 
\label{mass}
M(r)= \frac{1}{4} \int_{0}^{r} dr f^{-1} F^{\prime} =4\pi\int_{0}^{r} dr \rho f^{-1} R^{\prime} R^{2},
\end{equation}
so that
\begin{equation}
 M^{\prime} (r) = \frac{dM}{dr} = 4\pi\rho f^{-1} R^{\prime}R^{2}.
\end{equation}
For $\rho > 0$ everywhere we have $F^{\prime} >0$ and
$R^{\prime} >0$, so that in the non-singular part of the model $R>0$
except for $r=0$ and $F(r)$ is non-negative and monotonically
increasing for $r \geq 0$. 

 In a spherically symmetric inhomogeneous void model, we have two
``Hubble parameters'': $H_{\parallel}=H_{r}(t,r)$ for the local expansion rate in the radial
direction and $H_{\bot}(t,r)$ for expansion in the perpendicular direction~\cite{Moffat1,Moffat2}:
\begin{equation}
\label{hubble}
H_{\parallel}\equiv H_{r}=\frac{\dot{l}_{r}}{l_{r}}=\frac{\dot{R}^{\prime}}{R^{\prime}},
\end{equation}
\begin{equation} 
\label{hubblep}
H_{\bot}=\frac{\dot{l}_{\bot}}{l_{\bot}}=\frac{\dot{R}}{R},
\end{equation}
where $l$ denotes the proper distance, i.e. $dl_{r} = R^{\prime} (t,r) f^{-1}
dr$ and $dl_{\bot} = R(t,r) d\Omega$. Due to the fact that there are both
gravitational and expansion redshifts contributing to the total $z$, neither of
the Hubble parameters $H_{\parallel}$, $H_{\bot}$ is fully analogous to the LFRW's
$H_{LFRW}=\dot{a}/a$. For small $z$ we have
\begin{equation}
\label{small z}
 z (t_{e},r_{e})=H_{\bot}(t_{e},r_{e}) d_{L} (t_{e},r_{e}),
\end{equation}
where $t_e$ and $r_e$ denote the time and position emission of a light ray, respectively, and $d_L$ is the luminosity distance:
\begin{equation}
d_L\equiv\biggl(\frac{{\cal L}}{4\pi{\cal F}}\biggr)=R(t_e,r_e)[1+z(t_e,r_e)]^2,
\end{equation}
where ${\cal L}$ and ${\cal F}$ denote the luminosity and flux measured by an observer near $r=0$.  Eq.(\ref{small z}) is formally analogous to the LFRW result. Two main differences are that our relation is {\em local} and that from cosmological observations, 
we obtain the angular Hubble parameter $H_{\bot}=\dot{R} /R$ rather than $H_{LFRW}=\dot{a}/a$.

A test of the cosmological Copernican principle and the FLRW model is that for the FLRW model we have
\begin{equation}
{\cal H}(z)=H_{\parallel}-H_{\bot}(z)=0.
\end{equation}

\section{Observational Tests of Inhomogeneous and FLRW Cosmologies}

A critical test of an inhomogeneous model, such as the void cosmological model is whether the model can fit the CMB data and baryon acoustic oscillation (BAO) data. The simple spherically symmetric void cosmology with a homogeneous bang time and without a radiation dominant phase can fit the angular power spectrum data at the price of having a too low Hubble parameter $H_0$~\cite{Zibin}. The constraint on $H_0$ is lifted with a radially varying bang time $t_B$~\cite{Zuntz,Biswas}.
An alternative scenario for a void cosmology is to include a radiation phase in the model~\cite{ClarksonRegis1,ClarksonRegisLim}. Regis and Clarkson introduce a two-fluid model describing the matter and radiation components. They consider the Gaussian case:
\begin{equation}
\Omega_m(r)=\Omega_m^{(\rm out)}-(\Omega_m^{(\rm out)}-\Omega_m^{(\rm in)})\exp(-r^2/2\sigma^2)
=1-\Omega_k(r)-\Omega_r(r),
\end{equation}
where $\Omega_m,\Omega_k$ and $\Omega_r$ denote the matter, curvature and radiation profiles, respectively, and $\sigma$ is a constant. The $\Omega$ parameter values labeled (in) and (out) relate to the local and asymptotic $\Omega$ and other parameter values, respectively. The model is fully fixed by the parameters:
\begin{equation}
T_0,h,\Omega_m, f_b, \eta, N_{\rm eff},
\end{equation}
where $T_0,h,f_b,\eta$ and $N_{\rm eff}$ denote the CMB temperature today, the dimensionless Hubble rate, baryon fraction, and the baryon-photon ratio, and the number of relativistic degrees of freedom, respectively. We have
\begin{equation}
T_0^{(\rm in)}\sim 2.725\,K\quad {\rm and}\quad \Omega_\gamma^{(\rm in)}h^{(\rm in)2}\sim 2.469\times 10^{-5},
\end{equation}
which gives $\Omega_r^{(\rm in)}\sim 1.69\Omega_\gamma^{(\rm in)}$ for $N_{\rm eff}=3.04$.

Once a void model is fully specified, the power spectrum can be calculated. Regis and Clarkson~\cite{ClarksonRegis1} have fitted the CMB angular power spectrum by choosing the CMB shift parameters which characterize the key features of the first three power spectrum peaks~\cite{HuDodelson,Wang}:
\begin{equation}
l_a=\pi\frac{d_A(z_*)}{a_*r_s(a_*)},\quad l_{\rm eq}=k_{\rm eq}\frac{d_A(z_*)}{a_*}=H_{\rm eq}\frac{T_*}{T_{\rm eq}}d_A(z_*),\quad R_*=\frac{3\rho_b}{4\rho_\gamma}\biggl |_*
=\frac{3\Omega_b}{4\Omega_\gamma}a_*,
\end{equation}
where an asterisk denotes decoupling and $d_A$ denotes the angular luminosity distance. With the exception of $d_A(z_*)$, all quantities are local to the surface of the observed CMB. For the values
\begin{equation}
l_a=302,\quad l_{\rm eq}=136.6,\quad R_*=0.63, \quad f_b=0.17,\quad \eta_{10}=6.2,
\end{equation}
the choices $\Omega_m^{(\rm in)}\sim 0.25$ (which fits the SNIa) and $\Omega_m^{(\rm out)}\sim 0.5$ (open FLRW background) yield $h^{(in)}=0.73$ and  $h^{(\rm out)}= 0.67$~\cite{ClarksonRegis1,ClarksonRegisLim}. 

A void cosmology can explain the discrepancy between the locally measured value of the Hubble constant $H_0=73.24\pm 1.74\,{\rm km}\,{\rm s}^{-1}\,{\rm Mpc}^{-1}$ and the $H_0=66.93\pm 0.62\,{\rm km}\,{\rm s}^{-1}\,{\rm Mpc}^{-1}$ determined by the Planck data and the $\Lambda CDM$ model with 3 neutrino flavors having a mass of 0.06 eV. The local value of $H_0$ is higher than the $\Lambda CDM$ model value by 3.4$\sigma$~\cite{Riess11,Riess16,Planck2,Planck3,BOSS}. There is no problem making a void cosmology fit the area distance and the CMB power spectrum with $h^{(\rm in)}=0.73$ and $h^{(\rm out)}=0.67$, provided the model is extended beyond the simple dust model to include a radiation dominated early period and/or the big bang time $t_B$ is made inhomogeneous~\cite{BustiClarkson,BustiClarksonSeikel}.

One of the most important probes of the first instants of the post big bang is that the lightest nuclei were synthesized in observable abundances~\cite{Steigman}. In the $\Lambda CDM$ model, the observed abundances are in good agreement with the model, except for the big bang nucleosynthesis (BBN) probe of $^{7}{\rm Li}$~\cite{Cyburt}. The measured value of $^{7}{\rm Li}$ at $z=0$ and the value for $^{7}{\rm Li}$ derived from from the CMB and the $\Lambda CDM$ model disagree by $\sim 5\sigma$~\cite{Cyburt}. In the void model an ${\cal O}(1)$ difference in the baryon density inside and outside the void produces an ${\cal O}(1)$ difference in the spatial profile of the baryon-photon ratio $\eta$ at BBN. The SNIa observations imply a decrease in the local matter density -- the same as the measurements of the $^{7}{\rm Li}$ observations. In this way, the void cosmology can explain the $^{7}{\rm Li}$ problem if $\eta_{\rm CMB}=\eta^{(\rm out)}\sim1.5\times\eta^{(\rm in)}$ and at the same time explain the cosmological data without dark energy~\cite{RegisClarkson}.

A local void associated with large bulk flow generates CMB anisotropies connected to the kSZ effect. Let us suppose that the amplitude of the dipole induced by the radial bulk velocity is the dominant anisotropy. The difference in the redshift between incoming and outgoing CMB photons can be estimated. The dipole anisotropy at clusters is
\begin{equation}
\frac{\Delta T}{T}\sim 1-\frac{R(\eta_*)}{a(\eta_*)},
\end{equation}
where $a(\eta_*)$ and $R(\eta_*)$ denote the scale factors of the background LFRW and void regions, respectively, when the incoming photons enter the void. For a sufficiently large void with $h^{(\rm in)}=0.73$, we can obtain $\Delta T/T\sim 10^{-4}$ compared with the Planck collaboration data result $\Delta T/T\sim 6\times 10^{-4}$~\cite{Planck2,Planck3,Ichiki}, if we choose an inhomogeneous big bang time $t_B$~\cite{Bull,Zuntz}. Further weakening of kSZ anisotropic effects occur when we take into account $f_b(r)$ and $\eta(r)$ in the void model~\cite{Clarkson}.

The weak lensing parameter $A_{\rm lens}$ is defined as a scaling parameter affecting the lensing potential power spectrum, $C_l^{\rm lens}\rightarrow A_{\rm lens}C_l$, and the standard $\Lambda$CDM model has $A_{\rm lens}=1$. This means that the lensing of background galaxies, CMB, or any field of photons, can be influenced by the inhomogeneous void model modification parameter $\beta$. We can parameterize the lensing parameter $A_{\rm lens}$ by
\begin{equation}
A^{(\rm void)}_{\rm lens}=(1+\beta)A^{\rm LFRW}_{\rm lens},
\end{equation}
where $A^{(\rm void)}_{\rm lens}$ and $A^{\rm LFRW}_{\rm lens}$ denote the void weak lensing parameter and the standard $\Lambda$CDM model weak lensing parameter, respectively. We can write the angular power spectrum of lensing convergence $\kappa_c$ as~\cite{Smith,Caldwell,Melchiorri}:
\begin{equation}
C^{\kappa_c}_l=\frac{2}{\pi}\int k^2dk[I^{\kappa_c}(k)]^2P(k),
\end{equation}
where
\begin{equation}
I^{\kappa_c}(k)=\int d\chi g(\chi)[2(1+\beta)j_l(k\chi)]
\end{equation}
and $j_l(k\chi)$ are the spherical Bessel functions.
The range of values for $\beta$~\cite{Planck2,Planck3,Silk2} is $0.2\leq\beta\leq 0.35$, giving the lensing parameter the range of values:
\begin{equation}
1.2\leq A^{(\rm void)}_{\rm lens}\leq 1.35.
\end{equation}
This result is deduced from the constraint analysis corresponding to a void model correction to the standard $\Lambda$CDM model including weak lensing.  This means that the standard model $A_{\rm lens}$ anomaly disappears when the inhomogeneous void model is taken into account. More accurate data on the value of $A_{\rm lens}$ are needed to remove the possibility that the anomaly is due to small systematic experimental errors.

During the process of recombination, the baryons can become free from the photons and the size of the sound horizon at this time is imprinted as a bump in the two-point correlation function of the matter at late times. The baryon-acoustic-oscillation (BAO) data provide a measurement of the geometry of an inhomogeneous cosmological model, and in particular a probe of $H(z)$. For the simpler void models with zero bang time and no isocurvature modes, the BAO data are in tension with the SNIa data~\cite{Zibin,GarciaBellido}. As with the fits to the CMB data, the tension with the BAO data can easily be circumvented by allowing for an inhomogeneous bang time function, because this allows $H_{\parallel}$ to be fixed separately from $d_L(z)$. Alternatively, we can exploit the freedom in $f_b=f_b(r)$ and $\eta=\eta(r)$ to change $d_L(z)$ as a function of the radial shell about the observer~\cite{Clarkson,GarciaBellido2}.  In addition, there can be complications occurring from the evolution of perturbations that produce the tension in fitting the BAO data. 

The cluster abundance estimation can be related to the local cosmological parameters $H_{\rm loc}$ and $\Omega_{\rm loc}$ and the asymptotic background $H_{\rm bg}$ and $\Omega_{\rm bkg}$, which can be fixed by the relation:
\begin{equation}
H_{\rm loc}=\lambda H_{\rm bkg},
\end{equation}
where $\lambda > 1$. The parameter $\lambda$ has the effect of decreasing the cluster abundance with increasing $z$, caused by the smaller amplitude of density fluctuations due to a slower growth rate in the background LFRW spacetime. A fit to the local void prediction for the cluster number counts compared to the Planck SZ cluster number count data can be obtained for $\lambda=1.09$ for $h^{(\rm in)}=0.73$ and $h^{(\rm out)}=0.67$~\cite{Ichiki}.

\section{Conclusions} 

If we adopt an extended LTB void model including an epoch of radiation dominance and/or an inhomogeneous big bang time $t_B$, then as has been demonstrated
by Regis and Clarkson~\cite{ClarksonRegis1,ClarksonRegisLim}, it is possible to fit the data with a local $z=0$ value $h=0.73$ and a CMB value $h=0.67$. If the local measurements of $H_0$ at $z=0$ improve with future experiments, so that the measurement error reduces to 1\% or less, whereby the discrepancy between the local and CMB value of $H_0$ is $\ge 5\sigma$, then an inhomogenous cosmological model such as the void model can explain this significant observational anomaly in the FLRW standard model, and at the same time fit the CMB data. Moreover, the inhomogeneous model can resolve the lithium $^{7}{\rm Li}$ problem.  It can also resolve the CMB weak lensing anomaly and the anomalously low value of the local amplitude of density fluctuations compared with the prediction from the Planck data. In particular, it explains that the number of massive clusters obtained from the SZ effect is about half that expected from the CMB anisotropies.

The excellent fits of the CMB data by the standard FLRW model lead us to believe that it is the correct description of the large scale structure and evolution of the universe. Although an inhomogenous cosmological model such as the void model requires fine-tuning in space, the severe fine-tuning that occurs in time in the FLRW standard model, the tensions in the data, e.g., for $H_0$ and the lithium $^{7}{\rm Li}$ abundance, and the lack of understanding of the nature of dark energy require us to question how much confidence we should ascribe to the standard FLRW model~\cite{Buchert}. A rigorous test of the Copernican principle assumption and with it the FLRW scenario is needed. It may be possible to carry out a rigorous test of the Copernican principle over the coming years. Although the inhomogeneous cosmological model has more arbitrariness in its execution, the avoidance of a need to explain dark energy and the acceleration of the expansion of the universe through a cosmological constant with its fine-tuning problem, or from {\it ad hoc} modifications of general relativity, leads to a more conservative explanation for the large scale structure and evolution of the universe. 

\section*{Acknowledgments}

I thank Niayesh Afshordi, Alan Coley, Martin Green and Viktor Toth for helpful discussions. This research was supported in part by Perimeter Institute for Theoretical Physics. Research at Perimeter Institute is supported by the Government of Canada through the Department of Innovation, Science and Economic Development Canada and by the and by the Province of Ontario through the Ministry of Research, Innovation and Science.

\end{document}